\documentclass[preprint]{aastex}
\newcommand{\kms}{km~s$^{-1}$}
\usepackage{natbib}
\usepackage{graphicx}
\usepackage{amssymb}
\usepackage{txfonts}
\begin{document}

\title{Detection of the simplest sugar, glycolaldehyde, in a solar-type protostar with ALMA}
\author{Jes K. J{\o}rgensen\altaffilmark{1,2}, C\'ecile
  Favre\altaffilmark{3}, Suzanne E. Bisschop\altaffilmark{2,1}, Tyler
  L. Bourke\altaffilmark{4}, Ewine F. van Dishoeck\altaffilmark{5,6},
  and Markus Schmalzl\altaffilmark{5}}

\altaffiltext{1}{Centre for Star and Planet Formation and Niels Bohr Institute,
  University of Copenhagen, Juliane Maries Vej 30, DK-2100 Copenhagen
  {\O}., Denmark, jeskj@nbi.dk}
\altaffiltext{2}{Centre for Star and Planet Formation and Natural History Museum of Denmark, University of Copenhagen, {\O}ster Voldgade 5--7, DK-1350 Copenhagen
  {K}., Denmark, suzanne@snm.ku.dk}
\altaffiltext{3}{Department of Physics and Astronomy, Aarhus University, Ny Munkegade, DK-8000, Aarhus C., Denmark, favre@phys.au.dk}
\altaffiltext{4}{Harvard-Smithsonian Center for Astrophysics, 60 Garden Street, Cambridge, MA 02138, USA, tbourke@cfa.harvard.edu}
\altaffiltext{5}{Leiden Observatory, Leiden University, PO Box 9513, NL-2300 RA
Leiden, The Netherlands, ewine@strw.leidenuniv.nl, schmalzl@strw.leidenuniv.nl}
\altaffiltext{6}{Max-Planck Institut f\"ur extraterrestrische
Physik, Giessenbachstrasse, D-85748 Garching, Germany}

\begin{abstract}
  Glycolaldehyde (HCOCH$_2$OH) is the simplest sugar and an important
  intermediate in the path toward forming more complex biologically
  relevant molecules. In this paper we present the first detection of
  13~transitions of glycolaldehyde around a solar-type young star,
  through Atacama Large Millimeter Array (ALMA) observations of the
  Class~0 protostellar binary IRAS~16293-2422 at 220~GHz
  (6~transitions) and 690~GHz (7~transitions). The glycolaldehyde
  lines have their origin in warm (200--300~K) gas close to the
  individual components of the binary. Glycolaldehyde co-exists with
  its isomer, methyl formate (HCOOCH$_3$), which is a factor 10--15
  more abundant toward the two sources. The data also show a tentative
  detection of ethylene glycol, the reduced alcohol of
  glycolaldehyde. In the 690~GHz data, the seven transitions predicted
  to have the highest optical depths based on modeling of the 220~GHz
  lines all show red-shifted absorption profiles toward one of the
  components in the binary (IRAS16293B) indicative of infall and
  emission at the systemic velocity offset from this by about 0.2$''$
  (25~AU). We discuss the constraints on the chemical formation of
  glycolaldehyde and other organic species -- in particular, in the
  context of laboratory experiments of photochemistry of
  methanol-containing ices. The relative abundances appear to be
  consistent with UV photochemistry of a CH$_3$OH--CO mixed ice that
  has undergone mild heating.  The order of magnitude increase in line
  density in these early ALMA data illustrate its huge potential to
  reveal the full chemical complexity associated with the formation of
  solar system analogs.
\end{abstract}


\maketitle
\clearpage
\section{Introduction}\label{introduction}
One of the most intriguing questions in studies of the chemistry of
the early solar system is whether, how, when and where complex organic
and potentially prebiotic molecules are formed. One of the key species
in this context is glycolaldehyde (HCOCH$_2$OH). It is the simplest
sugar and the first intermediate product in the formose reaction that
begins with formaldehyde (H$_2$CO) and leads to the (catalyzed)
formation of sugars and ultimately ribose, the backbone of RNA, under
early Earth conditions \citep[e.g.,][]{larralde95}. The presence of
glycolaldehyde is therefore an important indication that the processes
leading to biologically relevant molecules are taking place. However,
the mechanism responsible for its formation in space is still unclear
\citep[see, e.g.,][]{woods12}.

Glycolaldehyde has so-far been detected in two places in space --
toward the Galactic center source SgrB2(N) (\citealt{hollis00}; see
also \citealt{hollis01,hollis04,halfen06,requenatorres08}) and the
high-mass hot molecular core G31.41+0.31 \citep{beltran09}. Another
compelling related discovery is that of ethylene glycol
(``anti-freeze''; (CH$_2$OH)$_2$), the reduced alcohol variant of
glycolaldehyde, found also toward SgrB2(N) at comparable abundances
\citep{hollis02}. Searches for glycolaldehyde in comets have so-far
only resulted in upper limits, whereas ethylene glycol is detected
toward Hale-Bopp and found to be at least 5 times more abundant than
glycolaldehyde \citep{crovisier04}. Comparisons between these species
are therefore particularly interesting as their relative abundances
potentially provide strong constraints on their formation and the
chemical evolution from protostars to primitive solar system material.

The protostellar (Class~0) binary IRAS16293-2422 (IRAS16293 hereafter)
at a distance of 120~pc \citep{loinard08} has long been considered to
be the best low-mass protostellar testbed for astrochemical studies
\citep[see, e.g.,][]{blake94,vandishoeck95,ceccarelli00a,schoeier02},
a status that has been further bolstered by the detection of a wealth
complex organic molecules toward this source
\citep{cazaux03,caux11}. (Sub)millimeter wavelength interferometric
studies of its chemistry have revealed strong differentiation among
different species toward the two components in the binary (IRAS16293A
and IRAS16293B; \citealt{wootten89}) including complex organic
molecules
\citep{bottinelli04iras16293,hotcorepaper,bisschop08,iras16293sma}.

With the Atacama Large Millimeter/submillimeter Array (ALMA) beginning
operations a completely new opportunity has arisen for studies of the
astrochemistry of solar-type stars. ALMA provides high sensitivity for
faint lines, high spectral resolution which limits line confusion, and
high angular resolution making it possible to study young stars on
solar-system scales. In this letter we report the first potential
discoveries of glycolaldehyde and ethylene glycol in a solar-type
protostar from ALMA observations of IRAS~16293-2422.

\section{Observations}
IRAS~16293-2422 was observed on 2011 August 16--17 as part of the ALMA
Science Verification (SV) program in band 6 \citep[see
also][]{pineda12}. At the time of observations 16 antennae were
present in the array in a compact configuration resulting in a
synthesized beam size of 2.5\arcsec$\times$1.0\arcsec (PA =
92$^\circ$). The source was observed in a two-point mosaic with a
total integration time of 5.5~hrs. The observations contain one
spectral window with 3840 channels and a channel width of 61.0~kHz
(0.083~\kms) covering a bandwidth from 220.078~GHz to 220.313~GHz. The
resulting line RMS noise level is estimated to be
13~mJy~beam$^{-1}$~channel$^{-1}$ in off-source line free channels.

IRAS~16293-2422 was further observed in ALMA's band 9 on 2012 April
16--17 with 13 antennae in the array. For these observations a
seven-point mosaic was performed with the array in an extended
configuration resulting in a synthesized beam size of
0.29$''$$\times$0.18$''$ (PA = 113$^\circ$) covering the full spectral
ranges 686.5--692.2~GHz (except 688.35-688.50~GHz) and
703.2--705.1~GHz with an effective spectral resolution of 980~kHz
(0.4~\kms). In total about 9.2~hrs were spent on the mosaic resulting
in an RMS of 0.11~Jy~beam$^{-1}$~channel$^{-1}$.

For both datasets we followed the reduction in the CASA
cookbooks/scripts delivered together with the SV data. To test the
reliability of the SV data we compared the resulting band~6 spectra to
our large Submillimeter Array survey
\citep[SMA,][]{iras16293sma,bisschop08}. Excellent agreement is seen
between the SMA and ALMA spectra at the two continuum peaks with
fluxes for bright lines agreeing to better than 10--20\%.  This
illustrates that the relative calibration of the two arrays is good
but also, equally important, that the detected line emission is
compact in both beams: the larger SMA beam of 4$''$$\times$2.4$''$
does not pick up significant extended emission that may have been
missed by ALMA.

\section{Results}
\begin{figure}
  \resizebox{\hsize}{!}{\includegraphics{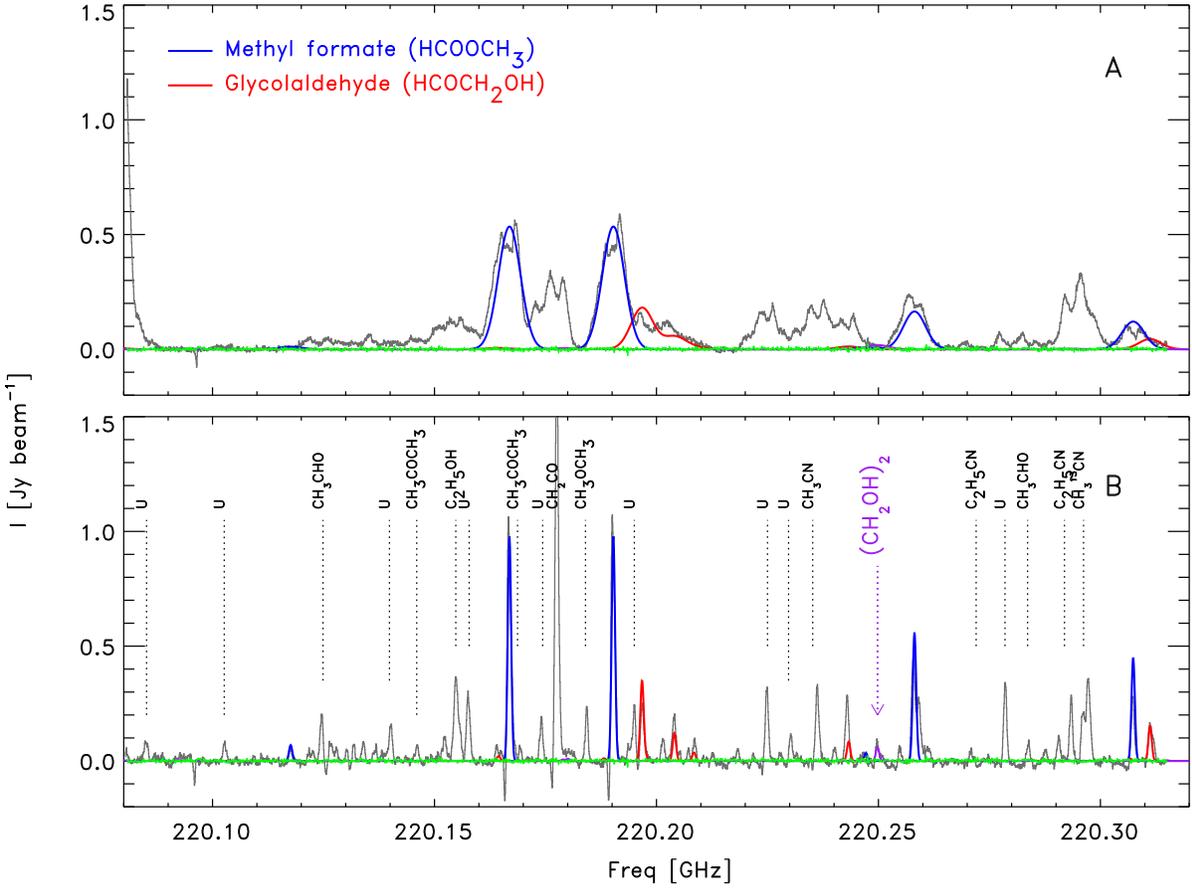}}
  \caption{Spectra in the central beams toward the continuum peaks of
    IRAS16293A (upper) and IRAS16293B (lower). Fits from LTE models of
    the methyl formate (blue) and glycolaldehyde (red) emission are
    overplotted. The purple line indicates the model fit to the
    possible ethylene glycol transition. The X-axis represents the
    frequencies in the rest frame of the system (i.e., corrected for
    the system $V_{\rm LSR}$ of 3~\kms). The green line is an
    indication of the RMS level (13~mJy~beam$^{-1}$) represented by a
    spectrum extracted from an off source position. Note the much
    narrower lines toward IRAS16293B which facilitate identification
    of individual features.}\label{spectrum}
\end{figure}

Fig.~\ref{spectrum} shows the band~6 spectra at 220.2~GHz within one
synthesised beam toward the two continuum peaks marking the locations
of IRAS16293A ($\alpha$=16\fh32\fm22.87\fs;
$\delta$=$-$24\fd28\arcmin36\farcs39) and IRAS16293B
($\alpha$=16\fh32\fm22.62\fs; $\delta$=$-$24\fd28\arcmin32\farcs46). A
large number of lines are clearly seen toward both positions. The
widths of the lines follow the pattern seen in previous studies, with
IRAS16293A showing lines about a factor of 5 broader than those toward
IRAS16293B. Also, as seen in previous observations
\citep[e.g.,][]{bottinelli04iras16293} many of the lines toward
IRAS16293B show red-shifted absorption features against the continuum
indicative of infalling motions.

\begin{figure}
\resizebox{0.7\hsize}{!}{\includegraphics{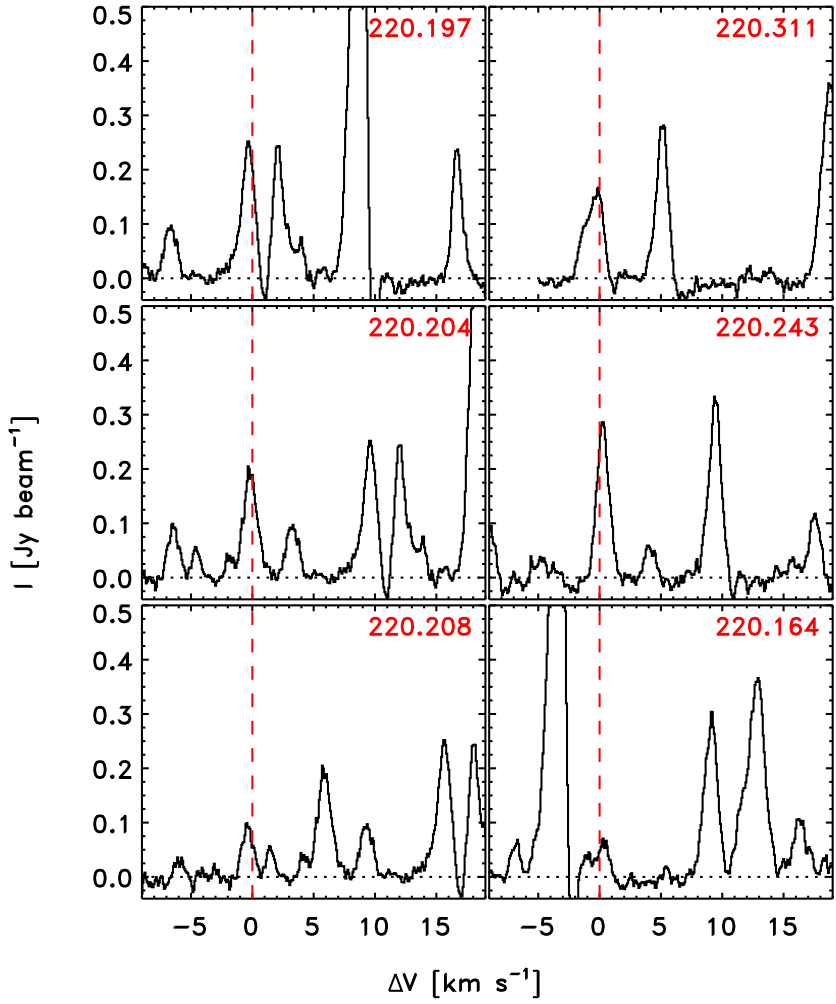}}
\caption{Zoom-in on the glycolaldehyde transitions detected at
  220~GHz. The velocities on the X-axis is given relative to the
  systemic velocity of 2.7~km~s$^{-1}$.}\label{band6}
\end{figure}

In the 220~GHz data the most easily identifiable species is methyl
formate, which is responsible for the two brightest lines at 220.1669
and 220.1903~GHz as well as a handful of other lines. A number of
other species are easily detected, including ketene (CH$_2$CO) at
220.1776~GHz and trans-ethanol (t-C$_2$H$_5$OH) at 220.1548~GHz
\citep[see also][]{bisschop08}. More remarkable is a set of clearly
detected lines that can be attributed to
glycolaldehyde. Table~\ref{line_id} lists the identified transitions
of methyl formate and glycolaldehyde, including lines in vibrationally
excited levels, together with references to the laboratory
spectroscopy data on which the frequencies are based.  Simultaneous
Gaussian fits to all lines in the central beams toward each of the
sources were made to determine line intensities. For IRAS16293A the
average width of all lines from the Gaussian fits is 6.3~\kms\ while
for IRAS16293B it is 1.4$\pm 0.3$~\kms, indicating very small
differences in widths between the lines. Likewise the average $V_{\rm
  LSR}$ is 3.0~\kms\ for IRAS16293A and 2.7~\kms\ for IRAS16293B with
standard deviations of only 0.1--0.2~km~s$^{-1}$. For comparison the
quoted uncertainties for the predicted line rest frequencies
correspond to 0.01--0.02~km~s$^{-1}$. The systemic velocities and line
widths are in good agreement with the large set of lines detected with
the SMA \citep{iras16293sma}. All lines of methyl formate and
glycolaldehyde have similar spatial extents toward each of the two
sources. Fig.~\ref{band6} shows close-ups of each identified
glycolaldehyde line toward IRAS16293B at 220~GHz.

\begin{table*}\centering
\caption{Identified lines and results from Gaussian fit to emission.}\label{line_id}
\begin{tabular}{llllllll}\hline\hline
Molecule      & Transition   & Frequency  & $\log_{10} A_{\rm ul}$ &
$E_u$      & \multicolumn{2}{c}{Flux [Jy~\kms]$^a$} & $V_{\rm LSR,B}$$^c$  \\
              &              &    [GHz]   & [s$^{-1}$]             &
              [K]        & $I_A$ & $I_B$ & [km~s$^{-1}$] \\ \hline
\multicolumn{8}{c}{Band 6 lines} \\ \hline
HCOOCH$_3$    & $33_{9,25}-33_{8,26}$A $\varv_t=1$     &  220.1176  & $-$4.18 & 572.6  & 0.14     &  0.063 & 2.5   \\
              & $17_{4,13}-16_{4,12}$E $\varv_t=0$     &  220.1669  & $-$3.82 & 103.2 & 4.36      &  1.39  & 2.6   \\
              & $17_{4,13}-16_{4,12}$A $\varv_t=0$     &  220.1903  & $-$3.82 & 103.1 & 4.16      &  1.43  & 2.6   \\
              & $24_{1,23}-24_{0,24}$E $\varv_t=1$     &  220.2472  & $-$5.46 & 355.1 & 0.056    &  0.061  & 2.9  \\
              & $18_{8,10}-17_{8,9}$E $\varv_t=1$      &  220.2581$^b$ & $-$3.89 & 330.8 & 1.89      &  0.66 & 2.7    \\
              & $24_{2,23}-24_{1,24}$E $\varv_t=1$     &  220.2585$^b$ & $-$5.46 & 355.1 & $\ldots$  &  $\ldots$ & $\ldots$ \\
              & $18_{10,9}-17_{10,8}$E $\varv_t=1$     &  220.3074  & $-$3.95 & 354.3 & 0.69      &  0.38  & 2.6   \\
HCOCH$_2$OH   & $11_{3,8}-10_{2,9}$ $\varv=1$    &  220.1644     & $-$4.37 & 323.5 & $\ldots$  &  0.092 & 2.9   \\
              & $7_{6,2}-6_{5,1}$ $\varv = 0$    &  220.1966$^b$ & $-$3.60 &  37.4 & 1.04      &  0.34 & 2.8    \\
              & $7_{6,1}-6_{5,2}$ $\varv = 0$    &  220.1968$^b$ & $-$3.60 &  37.4 & $\ldots$  &  $\ldots$ & $\ldots$ \\
              & $11_{4,7}-10_{3,8}$ $\varv = 0$  &  220.2040     & $-$3.99 &  46.6 & 0.83      &  0.28 & 2.6    \\
              & $11_{4,7}-10_{3,8}$ $\varv=2$    &  220.2084     & $-$3.98 & 420.8 & 0.12      &  0.13 & 2.4    \\
              & $36_{10,27}-36_{9,28}$ $\varv=1$ &  220.2433     & $-$3.69 & 713.0 & 1.14      &  0.38 & 2.9    \\
              & $7_{6,2}-6_{5,1}$ $\varv=1$      &  220.3111$^b$ & $-$3.60 & 318.2 & 0.25      &  0.26 & 2.5    \\
              & $7_{6,1}-6_{5,2}$ $\varv=1$      &  220.3113$^b$ & $-$3.60 & 318.2 & $\ldots $ &  $\ldots$ & $\ldots$ \\ \hline
\multicolumn{8}{c}{Band 9 lines$^{d}$} \\ \hline
HCOCH$_2$OH   & $38_{11,28}-37_{10,27}$ $\varv = 0$        &  686.6517  & $-$2.53 & 488.1 & $\ldots $ &  $\ldots$ &  $\ldots$ \\
              & $30_{14,16/17}-29_{13,17/16}$ $\varv = 0$  &  687.0513  & $-$2.30 & 377.4 & $\ldots $ &  $\ldots$ &  $\ldots$ \\
              & $19_{19,0/1}-18_{18,1/0}$ $\varv = 0$      &  687.4445  & $-$1.99 & 324.6 & $\ldots $ &  $\ldots$ &  $\ldots$ \\
              & $35_{12,24}-34_{14,23}$ $\varv = 0$        &  689.4295  & $-$2.42 & 438.8 & $\ldots $ &  $\ldots$ &  $\ldots$ \\
              & $35_{12,23}-34_{14,24}$ $\varv = 0$        &  689.5776  & $-$2.42 & 438.8 & $\ldots $ &  $\ldots$ &  $\ldots$ \\
              & $28_{15,13/14}-27_{14,14/13}$ $\varv = 0$  &  689.8903  & $-$2.24 & 362.1 & $\ldots $ &  $\ldots$ &  $\ldots$ \\
              & $27_{16,11/12}-26_{15,12/11}$ $\varv = 0$  &  703.8607  & $-$2.17 & 365.3 & $\ldots $ &  $\ldots$ &  $\ldots$ \\
              & $42_{11,32}-41_{10,31}$ $\varv = 0$$^{e}$  &  704.8369  & $-$2.57 & 580.0 & $\ldots $ &  $\ldots$ &  $\ldots$ \\\hline
\end{tabular}

Notes: Molecular data are taken from the JPL and CDMS catalogs
\citep{jpl,cdms1,cdms2}. The glycolaldehyde molecular data are based on laboratory measurements by
\cite{butler01}, \cite{widicus05} and \cite{carroll10} and
model predictions based on these (see text) as provided by the JPL June 2012 catalog entry. $^a$Total
flux from Gaussian fits to the line emission in the central beam toward
IRAS16293A ($I_A$) and IRAS16293B ($I_B$). For conversion to brightness
temperatures, the gain of the interferometric observations with the
given beam size is 0.1~Jy~K$^{-1}$. $^b$Transitions not
resolved. $^c$$V_{\rm LSR}$ toward IRAS16293B. $^d$Lines show emission off-source and red-shifted
absorption on-source. $^{e}$Shown in
Fig.~\ref{band9}; not detected. 
\end{table*}

\begin{figure}
\resizebox{0.7\hsize}{!}{\includegraphics{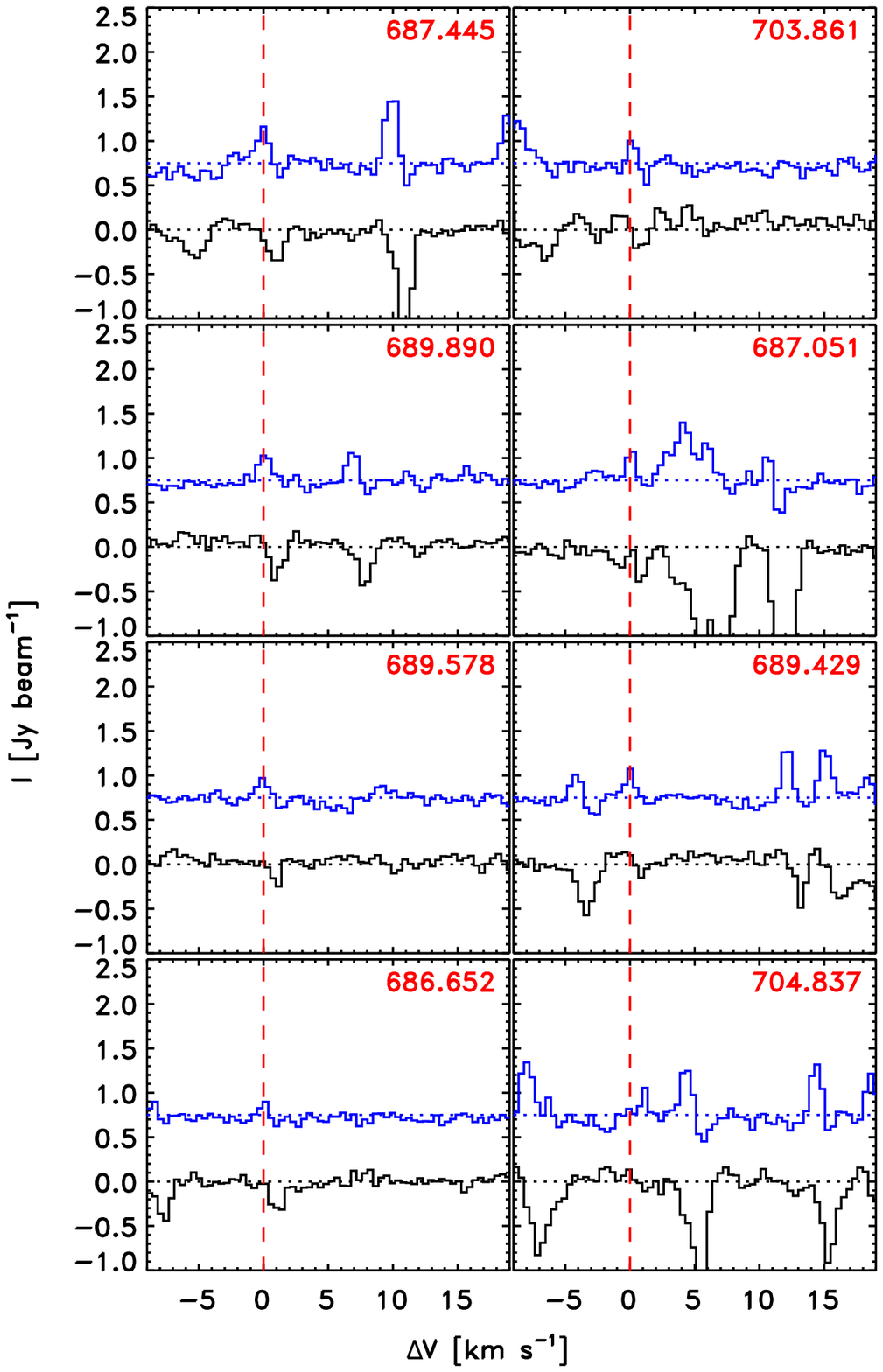}}
\caption{Zoom-in on the glycolaldehyde transitions predicted to be
  most optically thick at 690~GHz toward the IRAS16293B continuum peak
  (\emph{black}) and toward a position offset by one synthesized beam
  from the continuum peak (\emph{blue}; shifted by 0.75~Jy~beam$^{-1}$
  in the Y-axis direction). As in Fig.~\ref{band6} the velocities on
  the X-axis is given relative to the systemic velocity.}\label{band9}
\end{figure}

To further validate the detections of the methyl formate and
glycolaldehyde lines, we modeled the strengths for all their lines in
the observed band assuming an isothermal medium and LTE excitation and
taking into account the optical thickness of the lines \citep[see,
e.g.,][]{goldsmith99}. Fig.~\ref{spectrum} includes the best fit
models to the methyl formate and glycolaldehyde lines. The fits show
comparable temperatures for the two species (200 and 300~K for
IRAS16293A and IRAS16293B, respectively). The derived column densities
averaged over 0.5$''$ (60~AU) regions are given in
Table~\ref{modelfit}. The model reproduces the observed line profiles
and strengths very well, including the vibrationally excited lines,
even for this fairly simplified model.

The 220~GHz data also show a tentative detection of ethylene
glycol. The strongest line of ethylene glycol
($22_{2,20}\,(v=1)-21_{2,19}\,(\varv=0)$ at 220.2498~GHz; $E_u =
127$~K) in the band coincides with a $\approx$18$\sigma$ line toward
IRAS16293B that is difficult to attribute to any other species. Using
the LTE model at 300~K, the observed line strength would require an
ethylene glycol abundance of 0.3--0.5 with respect to glycolaldehyde.
This is comparable to the estimates by \cite{hollis02} for SgrB2(N),
but notably smaller than the inferred values from comet Hale-Bopp
where ethylene glycol is at least five times more abundant than
glycolaldehyde \citep{crovisier04}.

Similar model calculations can rule out some of the alternative
identifications for the lines ascribed to
glycolaldehyde. Cyclopropenylidene (c-C$_3$H$_2$) for example has two
transitions at 220.2025~GHz that potentially could be confused with
the glycolaldehyde line at 220.2040~GHz. However, these specific
transitions of c-C$_3$H$_2$ have very low Einstein A coefficients
($\sim 10^{-8}$~s$^{-1}$), so that for any reasonable excitation
temperature, other stronger lines of this species should have been
observed in the frequency range covered by the SMA observations --
e.g., between 230.5 and 231.0~GHz\footnote{Lines of this species
  detected previously using single-dish observations have
  significantly higher Einstein A coefficients,
  $\sim$$10^{-4}$--$10^{-3}$~s$^{-1}$, and are likely somewhat
  extended.}. Also, if ascribed to c-C$_3$H$_2$, this emission would
be red-shifted by about 2~\kms\, which should be easily discernable
toward IRAS16293B. Likewise the transitions at 220.196~GHz could also
be attributed to a vinyl cyanide (CH$_2$CHCN) line at
220.1964~GHz. However, this species would have a much stronger
transition at 220.138~GHz that is not seen.

A strong confirmation of the identification of glycolaldehyde is
provided by the 690~GHz observations. A large number of glycolaldehyde
transitions are expected to be located in this band with a range of
energy levels and line strengths. The model based on the 220~GHz data
predicts that a number of these transitions are becoming optically
thick ($\tau_{\rm 60~AU} \gtrsim 0.1-1.0$ averaged over the 0.5$''$
region used for the column density estimates). Fig.~\ref{band9} shows
the spectra toward the IRAS16293B continuum peak as well as offset by
one synthesized beam for the lines expected to be most optically
thick. Toward the continuum peak, the 7 most optically thick
transitions all show red-shifted absorption against the IRAS16293B
continuum whereas they are seen in emission at the offset
position. Again, these lines show a nearly perfect correspondance to
the expected systemic velocity of 2.7~km~s$^{-1}$. Equally important,
the models do not predict other lines of glycolaldehyde that should be
detectable elsewhere in the large 690 GHz spectral window at the
sensitivity of the observations, nor in the 220 GHz window.  The
absence of emission toward the continuum peak for IRAS16293B suggests
that the continuum is completely optically thick at these high
frequencies and scales (0.25$''$; 30~AU diameter). This is consistent
with the observations of \cite{chandler05} who showed that the
continuum at centimeter wavelengths was due to optically thick thermal
dust emission at the same scales (0.15--0.25$''$). The fact that
glycolaldehyde and other complex species show infall signatures toward
source B \citep[see also][]{pineda12} implies that these molecules are
moving toward the planet-forming zones at $\leq$ 30 AU radius.

The main uncertainty in the assignments is the accuracy and
completeness of the spectroscopic catalogs. About one third of the
lines brighter than about 0.1~Jy~beam$^{-1}$ remain unidentified in
the 220~GHz spectrum and other possible assignments for the potential
glycolaldehyde lines can thus not be ruled out. Also, not all
glycolaldehyde line frequencies have been measured directly in the
laboratory, especially for the vibrationally excited states detected
here and for the lines at 690 GHz, but are based on a spectroscopic
model using measurements at other frequencies. However, since the
laboratory data cover lines up to 1.2~THz and include other
vibrationally excited lines with a similar range of $J$ and $K_a$
values as observed here, the frequency predictions should be
reliable. Indeed, their quoted uncertainties are lower than what can
be discerned in our data. A check can be made by comparing the
frequencies from the JPL and CDMS catalogs which have been computed
using slightly different methods and datasets. For the lines detected
in the ALMA spectra, the agreement between the frequencies is about
0.2~MHz -- corresponding to $<$~0.1~\kms. Thus, within the
uncertainties given in the catalogs the identification of lines are
secure. Also, the relative column density (or abundance) of methyl
formate to glycolaldehyde of 10--15 is consistent with previous
measurements (see also \S~\ref{chemistry}) and, combined with the
additional high frequency detections and the close matches in velocity
for 13 lines, provide a compelling case for the assignments. We stress
that IRAS16293B with its narrow line widths provides a much cleaner
source for identification of complex species than high mass sources
like SgrB2(N) that have been studied so far.  Future searches with
ALMA will be able to further strengthen the assignments either by
extending the number of lines or searching for specific spectroscopic
signatures -- e.g., those provided by the low excitation $\varv=0$
transitions used for previous detections at 3~mm.

\section{Discussion: formation of glycolaldehyde}\label{chemistry}
The similar spectral shapes of the lines of the complex organic
molecules (in particular, the line widths and the absorption profiles
toward IRAS16293B) as well as their similar spatial extent suggest
that methyl formate, glycolaldehyde and other species coexist in the
same gas. Furthermore, methyl formate and glycolaldehyde are fit by
similar excitation temperatures in our simple LTE models.  Therefore,
a discussion of the chemical relation between glycolaldehyde and
methyl formate is relevant.

Our results indicate that methyl formate is a factor 10--15 more
abundant than glycolaldehyde in the warm gas toward the two binary
components of IRAS16293-2422. This value is consistent with previous
measurements ranging from 52 in the hot core of SgrB2(N)
\citep{hollis01} and the upper limit of 34 in G34.41+0.31
\citep{beltran09} to the average of 5--6.5 found on more extended
scales toward SgrB2(N) \citep{hollis00,requenatorres08}. Relative to
the lower limit on the H$_2$ column density from the optically thick
dust continuum emission toward IRAS16293B \citep{chandler05} and
taking into account the filling factor, the two species are estimated
to have abundances relative to H$_2$ of 8$\times 10^{-8}$ (methyl
formate) and 6$\times 10^{-9}$ (glycolaldehyde).

One of the major discussions concerning the origin of complex organic
molecules in space is whether these form from second generation
gas-phase reactions based on protonated CH$_3$OH (released from ices
at high temperatures) or due to first generation reactions in the icy
grain mantles, possibly induced by UV- or cosmic-ray irradiation
\citep[e.g.,][]{herbst09}. Grain-surface reactions are becoming
increasingly more popular, especially since the formation of methyl
formate through gas-phase reactions seems to be too inefficient to
explain the observed abundances \citep{horn04}.

\cite{halfen06} compared a survey of the glycolaldehyde emission
toward SgrB2(N) with formaldehyde (H$_2$CO), motivated by the first
step in the formose reaction consisting of two H$_2$CO molecules
combining to form HCOCH$_2$OH. Modeling the emission from
H$_2$C$^{18}$O transitions detected in our SMA survey with the same
excitation temperature as for methyl formate and glycolaldehyde
suggests a formaldehyde to glycolaldehyde abundance ratio of 42--56
(using a $^{16}$O:$^{18}$O abundance ratio of 560 characteristic for
the local ISM \citep{wilson94}). This is in agreement with the
estimate by \cite{halfen06} of a ratio of 27 in
SgrB2(N). \citeauthor{halfen06} interpreted this ratio in favor of the
formation of glycolaldehyde in space through a gas-phase formose
reaction. However, as pointed out by \cite{woods12} this and other
gas-phase reactions tend to produce too little glycolaldehyde compared
to observed abundances in an absolute sense.

Alternatively glycolaldehyde may be formed through grain-surface
reactions in ices rich in methanol (CH$_3$OH) or its derivatives. From
laboratory experiments \cite{oberg09} found that photochemistry of
UV-irradiated methanol ices mixed with significant amounts of CO can
explain the observed fractions of oxygen-rich complex organics like
methyl formate relative to methanol in sources such as IRAS16293
\citep{bisschop08}. Although \cite{oberg09} could not fully separate
glycolaldehyde and methyl formate in their methanol-rich experiments,
the derived lower limits on their abundances in CO-containing ices
relative to methanol are 4\% and 8\%, respectively, consistent with
the combined IRAS16293 results of \cite{bisschop08} and those found
here of $\approx 1$\% and 10--20\%. Also, their upper limit of
ethylene glycol relative to glycolaldehyde ($<$25\%) agrees roughly
with our ratio of 0.3--0.5. Experiments with irradiation of pure
CH$_3$OH ices on the other hand produce too large ethylene glycol
abundances relative to glycolaldehyde by an order of magnitude.  The
CH$_3$OH:CO ratio is clearly critical: laboratory results for the
CH$_3$OH:CO=1:10 mixtures show a 4 times larger abundance of
glycolaldehyde relative to ethanol, whereas the ethanol lines detected
here and in the previous SMA observations \citep{bisschop08} suggests
that ethanol is 3--5 times more abundant than glycolaldehyde in
IRAS16293. As pointed out by \cite{oberg09} the relative abundances of
some species also strongly depend on the ice temperature, with the
glycolaldehyde and ethylene glycol production requiring heating above
$\sim$30 K.  Thus, both exact ice composition (amount of CO mixed with
CH$_3$OH) and temperature play a role in the chemistry; a combination
of a moderately CO-rich ice and mild heating best reproduce the
current data.

Still, additional systematic surveys of more species and sources are
needed to constrain the surface formation mechanisms in more
detail. These early data illustrate the enormous potential of ALMA for
doing this.  The current sensitivity is already more than an order of
magnitude better than that of previous single-dish or interferometric
line surveys toward this source \citep{caux11,iras16293sma}, revealing
a line density toward IRAS16293B nearly ten times higher than
before. Clearly, ALMA is posed to reveal many more complex organic
molecules in young solar-system analogs.

\acknowledgments 

The authors are grateful to Holger M\"{u}ller and an anonymous referee
for good discussions about the spectroscopic accuracy and useful
comments on the paper. This paper makes use of the following ALMA
Science Verification data:
\dataset{ADS/JAO.ALMA\#2011.0.00007.SV}. ALMA is a partnership of ESO
(representing its member states), NSF (USA) and NINS (Japan), together
with NRC (Canada) and NSC and ASIAA (Taiwan), in cooperation with the
Republic of Chile. The Joint ALMA Observatory is operated by ESO,
AUI/NRAO and NAOJ.  This research was supported by a Lundbeck
Foundation Junior Group Leader Fellowship and by a grant from the
Instrumentcenter for Danish Astrophysics (IDA) to Jes
J{\o}rgensen. Research at Centre for Star and Planet Formation is
funded by the Danish National Research Foundation and the University
of Copenhagen's programme of excellence. Astrochemistry in Leiden is
supported by a Spinoza grant from the Netherlands Organization for
Scientific Research (NWO), the Netherlands Research School in
Astronomy (NOVA), and by EU A-ERC grant 291141 CHEMPLAN. Markus
Schmalzl is supported by the ALMA Regional Center node Allegro funded
by NWO.

\begin{table}\centering
  \caption{Results of the model fits to the methyl formate and glycolaldehyde.}\label{modelfit}
\begin{tabular}{lll}\hline\hline
                              & I16293A  & I16293B \\ \hline
Temperature                   & 200~K    & 300~K   \\
$N_{\rm 60 AU}$(Methyl formate)           & 5$\times 10^{17}$~cm$^{-2}$ & 4$\times 10^{17}$~cm$^{-2}$        \\
$N_{\rm 60 AU}$(Glycolaldehyde)           & 4$\times 10^{16}$~cm$^{-2}$ & 3$\times 10^{16}$~cm$^{-2}$        \\ \hline
\end{tabular}
\end{table}

\end{document}